\documentclass[twoside,english,nofootinbib,preprintnumbers,aps,onecolumn,notitlepage,11pt]{revtex4}
\usepackage{euscript,amssymb}
\usepackage{amsthm}
\usepackage{graphicx}
\usepackage{amsfonts}
\usepackage{amsmath}
\usepackage{amssymb}
\usepackage{fancyhdr}
\usepackage{epsfig}
\usepackage{color}
\usepackage{soul}
\usepackage{calc}
\usepackage{enumerate}
\usepackage{afterpage}
\usepackage[papersize={8.5in,11in}]{geometry}

\allowdisplaybreaks

\newcommand{\be}{\begin{equation}}
\newcommand{\ee}{\end{equation}}

\begin{document}

\title{The Near Horizon Geometry Equation on Compact 2-Manifolds Including the General Solution for $g>0$.}
\author{Denis Dobkowski-Ry{\l}ko}
	\email{Denis.Dobkowski-Rylko@fuw.edu.pl}
	\affiliation{Faculty of Physics, University of Warsaw, ul. Pasteura 5, 02-093 Warsaw, Poland}
	\author{Wojciech Kami\'nski}
	\email{Wojciech.Kaminski@fuw.edu.pl}
	\affiliation{Faculty of Physics, University of Warsaw, ul. Pasteura 5, 02-093 Warsaw, Poland}

\author{Jerzy Lewandowski}
	\email{Jerzy.Lewandowski@fuw.edu.pl}
	\affiliation{Faculty of Physics, University of Warsaw, ul. Pasteura 5, 02-093 Warsaw, Poland}
\author{Adam Szereszewski}
	\email{Adam.Szereszewski@fuw.edu.pl}
	\affiliation{Faculty of Physics, University of Warsaw, ul. Pasteura 5, 02-093 Warsaw, Poland}
\begin{abstract} 
The Near Horizon Geometry (NHG) equation with a cosmological constant $\Lambda$ is considered on compact
$2$-dimensional manifolds. It is shown that every solution satisfies  the Type D equation at every point
of the manifold. A similar result known in the literature was valid only for non-degenerate 
in a suitable way points of a given solution. At the degenerate points the Type D equation was not applicable.  
In the current paper we prove that the degeneracy is ruled out by the compactness. Using that result we  find 
all the  solutions to the NHG equation on compact $2$-dimensional manifolds  of non-positive Euler characteristics. Some integrability
conditions known earlier in the  $\Lambda=0$ case are generalized to arbitrary value of $\Lambda$. They may be
still useful for compact $2$-manifolds of positive Euler characteristic.      
\end{abstract}

\date{\today}

\pacs{???}

\maketitle
\section{Introduction} 
The Near Horizon Geometry Equation is defined on a manifold $S$ endowed with a metric tensor $g_{AB}$,
and a differential $1$-form $\omega_A$, namely  it reads
 \begin{equation}\label{NHGnd}
 \nabla_{(A}\omega_{B)} + \omega_A\omega_B - \frac{1}{2}R_{AB} + \frac{1}{2}\Lambda g_{AB}=0
 \end{equation}
where  $\nabla_A$ is the torsion free covariant derivative defined on $S$ by $g_{AB}$,
that is
 \begin{equation}\label{nabla}
 \nabla_C g_{AB}=0 = (\nabla_A\nabla_B -\nabla_B\nabla_A)f,\ \ \ {\rm for \ every} \ \ \ f\in C^2(S), \end{equation}
 $R_{AB}$ is the Ricci tensor of $g_{AB}$ and  $\Lambda$ is a real constant. When $S$ is a $(n-2)$-dimensional 
 spacelike section of  an extremal  isolated horizon (for example a degenerate Killing horizon) in $n$-dimensional 
 spacetime  that satisfies the Einstein equations  with the cosmological constant, then $g_{AB}$ is the induced metric tensor, 
 $\omega_A$ is the induced rotation $1$-form  potential, and  $\Lambda$ is the cosmological constant 
 \cite{ABL1,Hajicek,IM,LPhigh}.   On the other hand, given any solution  $(g_{AB},\omega_A)$ defined on some $S$,
 there is a construction of an explicit ("exact")    solution to the Einstein equations defined on 
 $S\times\mathbb{R}\times\mathbb{R}$ \cite{PLJ, Kunduri1,Kunduri2}.  The first examples of those spacetimes were obtained        
 by the Bardeen-Horowitz limit from neighborhoods of the extremal horizons in  Kerr spacetimes - that is where
 the name  Near Horizon Geometry have come from \cite{Horowitz,CRT2}.      

The NHG equation attracts attention of researchers for several reasons.  The first one is scientific curiosity of whether there
may be solutions different, than embeddable in the extremal Kerr spacetimes  \cite{LPextremal}.
Secondly, every new solution to the NHG equation  would automatically lead to a new NHG solution to the Einstein equations. 
Thirdly,  the knowledge of all possible  degenerate Killing horizons is  important for filling gaps in the black hole  uniqueness theorems \cite{CCH}.   

In the current paper we consider the NHG equation on  $2$-dimensional manifolds $S$, hence it corresponds to the spacetime dimension $4$.    
The equation still has some secrets. The axisymmetric solutions on topological sphere $S_2$ with $\Lambda=0$ form a $1$-dimensional family 
that can be parametrized  by the area \cite{LPextremal} (see also \cite{Hajicek}). All of them correspond to the (spacelike sections) of horizons in extremal Kerr spacetimes.  That result was generalized to $\Lambda\not=0$ \cite{Kunduri1,Kunduri2}. What is  not known, is whether non-symmetric solutions exist or not on $S_2$. There seem to be none in a neighborhood of the axisymmetric ones \cite{CST}, and definitely there are no static solutions on $S_2$ \cite{CRT1}. The topological constraint \cite{PLJ,JK}  restricts possible genus and values of $\Lambda$. We discuss it in detail below. For genus and values of $\Lambda$ allowed by the topological constraint, some partial results are known in the literature: no axisymmetric solutions on torus \cite{Li}, all the static solutions on arbitrary compact $S$ have  $\omega_A=0$ and constant Ricci scalar \cite{CRT1}.
The local properties of the equation were studied in \cite{J,JK,NR}.          

Recently new insights on the NHG equation came from the study of the Type D equation \cite{DLP1, DKLS} that has the same unknowns, 
namely $g_{AB}$ and $\omega_A$ defined on a $2$-dimensional manifold $S$. The Type D equation  is satisfied by every solution 
to the NHG equation at every non-degenerate  point of $S$, that is  such that  the Gauss curvature of $g_{AB}$  is not equal to $\frac{\Lambda}{3}$ or $d\omega$ is not zero.  The main technical result of the current paper is a proof that  for a compact $S$ solutions of the NHG equation do not
have degenerate points except for the case that every point in $S$ is degenerate. The latter solution is easy to identify as a flat metric $g_{AB}$, 
identically zero $\omega_A$ and $\Lambda=0$.  Our proof opens the door to the application of the Type D equation which all  solutions on genus$(S)>0$ are known \cite{DKLS}.

\section{The equation and the topological constraint for $2$-dimensional $S$.}
In this paper we consider the NHG equation on a $2$-dimensional compact manifold $S$. 
For simplicity let us assume $S$ is connected and orientable. The generalization of our results
to the non-connected case is obvious, while generalization to non-orientable surfaces turns out to be easy as well
and will be discussed in the last section.   

In $2$ dimensions 
 \begin{equation}
 R_{AB}  =Kg_{AB},
 \end{equation}
 where $K$ is called the Gauss curvature. Hence, the NHG equation (\ref{NHGnd}) we study in this paper reads
\cite{ABL1}
 \begin{equation}\label{NHG2d}
 \nabla_{(A}\omega_{B)} + \omega_A\omega_B - \frac{1}{2}Kg_{AB} + \frac{1}{2}\Lambda g_{AB}=0.
 \end{equation}
The contraction with $g^{AB}$ and integration along $S$ with the area $2$-form  $\eta_{AB}$  corresponding to the metric $g_{AB}$ 
gives the following topological constraint \cite{PLJ,J} 
\begin{align}
 \frac{4\pi}{A} (1-{\rm genus})  \ =&\ \frac{1}{A}\int_S\omega_A\omega^A \eta\ +\Lambda \geq \ \Lambda\\
A:=&\int_S  \eta
 \end{align}
That  constraint allows for:  
\begin{enumerate}
\item all the values of $\Lambda$, for  the genus$(S)=0$ (sphere);
\item only $\Lambda\leq 0$, for genus$(S) =1$ (torus), in particular for $\Lambda=0$ the only solutions are  
flat $g_{AB}$ and $\omega_A=0$;
\item  only $\Lambda<0$, for genus$(S) >1$. 
\end{enumerate} 

In case 1  the family of axisymmetric solutions is known, whereas the existence of other solutions is an open problem
\cite{LPextremal,CST}.  In case 2 the non-existence of axisymmetric solutions of $\Lambda<0$ \cite{Li} is known.  In case 3, it is 
known \cite{CRT1}, that all the static solutions, namely
 \begin{equation}
d\omega=0 
 \end{equation}
 have a constant Gauss curvature
  \begin{equation}
K={\rm const}.
 \end{equation}
In this paper we will complete the solution of the NHG equation (\ref{NHG2d}) in case 3.  

\section{The complex valued scalar}
An important role in the study of the NHG equation is played 
by the following complex valued (almost) scalar \cite{ABL1,LPextremal}
\begin{equation}\label{Psi2}
\Psi_2\ :=\ \-\frac{1}{2}\left(K-\frac{\Lambda}{3}+i\Omega \right)
\end{equation}
where $\Omega$ is the pseudo scalar characterizing the rotation $2$-form, namely
 \begin{equation}\label{Omega}
\Omega\, \eta_{AB} \ :=\   2\nabla_{[A}\omega_{B]}  . 
\end{equation}
From now on, it will be convenient to introduce a null frame $m_A$ such that
\begin{equation}\label{mm}
g_{AB}\ =\ m_A\bar{m}_B +  m_B\bar{m}_A, \ \ \ \ \ \ \ \ \ \eta_{AB}\ =\ i\left(\bar{m}_Am_B- \bar{m}_Bm_A\right). 
\end{equation}
A null frame $m_A$ is defined locally on $S$, up to the transformations 
\begin{equation}\label{m'}
m'_{A} = e^{i\phi}m_A,  \end{equation}
where $\phi$ is a locally defined function. Denote by $m^A$ the dual frame, such that
\begin{equation}\label{m^A}\
m^Am_{A} = 0, \ \ \ \ \ m^A\bar{m}_A = 1 .
\end{equation}

The key observation is the following integrability condition for (\ref{NHG2d}):
\medskip

\noindent{\bf Proposition 1} {\it If a metric tensor $g_{AB}$ and a $1$-form $\omega_A$
defined on a $2$d manifold $S$ satisfy the NHG equation, then the scalar $\Psi_2$ satisfies 
the following equation}
\begin{equation}\label{deltaPsi}
\bar{m}^A\left(\nabla_A + 3\omega_A\right)\Psi_2\ =\ 0.  
\end{equation}  
\medskip

Equation (\ref{deltaPsi}) is invariant with respect to the transformations (\ref{m'}), that makes  it independent
 of the choice of $m^A$ and globally defined on $S$.  It can be derived by acting on (\ref{NHG2d}) with $\nabla_C$,
 and commuting. This equation was found in the $\Lambda=0$ case in \cite{LPextremal}, but it is true for arbitrary $\Lambda$
 with suitably defined $\Psi_2$. 
 \medskip
 
Equation (\ref{deltaPsi}) implies important properties of $\Psi_2$. The first one follows from the following lemma: 
\medskip 
  
 \noindent{\bf Lemma 1} {\it Suppose $S$ is a compact oriented $2$-manifold endowed with a metric tensor $g_{AB}$ and a $1$-form $\omega_A$.
 Suppose that a function $F:S\rightarrow \mathbb{C}$ satisfies the following equation (see (\ref{mm},\ref{m^A}))
  \begin{equation}\label{delta+3pi}
\bar{m}^A\left(\nabla_A + 3\omega_A\right)F\ =\ 0. 
\end{equation} 
Then either
 \begin{equation}
 F(x)\not= 0, \ \ \ \ \ {\rm for\ every}\ \ \ \ x\in S, 
\end{equation}
or 
\begin{equation}
 F(x) = 0, \ \ \ \ \ {\rm for\ every}\ \ \ \ x\in S. 
\end{equation}   
 }
\medskip

For $S=S_2$ (topologically), the proof can be found in   \cite{LPextremal} (see also \cite{CST}).  
The class of local frames $m^A$  related to each other by (\ref{m'}) allows to define uniquely 
a decomposition of the complexified tangent space 
$\mathbb{C}T_xS $ into the algebraic direct sum
\begin{align}
\mathbb{C}T_xS\ &=\ T_x^{(1,0)}\oplus T_x^{(0,1)}\\
\left(am^A + b\bar{m}^A\right)^{(1,0)} &= am^A \\
\left(am^A + b\bar{m}^A\right)^{(0,1)} &= b\bar{m}^A,
\end{align}
where $a$ and $b$ are arbitrary, complex valued, coefficients.
It is accompanied by the dual decomposition of the cotangent space
\begin{align}
\mathbb{C}T^*_xS\ &=\ T^*{}^{(1,0)}\oplus T^*{}^{(0,1)}\\
\left(am_A + b\bar{m}_A\right)^{(0,1)} &= am_A \\
\left(am^A + b\bar{m}^A\right)^{(1,0)} &= b\bar{m}_A.
\end{align}
The decompositions are invariant with respect to the transformations (\ref{m'}).

We will also use  complex valued  coordinates $(z,\bar{z})$ defined locally, in a neighborhood of
every point on $S$, such that 
\begin{equation} 
m^A\partial_A=P\partial_z, \ \ \ \ \ \bar{m}_A dx^A = \frac{1}{P}dz,
\end{equation}
with some locally defined  function $P$.  

With that notation,  the conclusion (\ref{deltaPsi}) reads
\begin{equation}
(\partial_z +3\omega_z) \Psi_2=0,
\end{equation}
where
$$
\omega=\omega_z dz + \omega_{\bar{z}}d{\bar z} = \omega^{(1,0)}+\omega^{(0,1)}.
$$
\medskip

\noindent{\it Proof of Lemma 1.}
 
Equation (\ref{delta+3pi})  now reads
\begin{equation}\label{delta+3pi'}
\partial_{\bar z} F+3\omega_{\bar z} F=0.
\end{equation}
 
In any simply connected open set $U\subset S$, there is a function $\phi_U$ such that
\begin{equation}
\partial_{\bar z}\phi_U=3\omega_{\bar z}. 
\end{equation}
The  eq. (\ref{delta+3pi}) implies
\begin{equation}
\partial_{\bar z} \left(F e^{\phi_U}\right)=0.
\end{equation}
 
Thus, the function
\begin{equation}\label{f_U}
f_U:=Fe^{\phi_U}
\end{equation}
is holomorphic in $U$.  It either does not vanish in $U$, has only isolated zeros in $U$ or is identically zero. 
In the latter case, by patching $S$ with such open sets,  we can prove that $F$ is identically zero on $S$. 
Hence $F$  either does not vanish in $S$ at all, is identically zero or has only isolated zeros in $S$.

Suppose  $F$ has  isolated zeros.  As our surface $S$ is compact there might be only 
finitely many zeros, denote them  $x_1,...,x_k\in S$.  Consider any of those points, $x_i$, say. Without the lack
of generality we  assume that the complex coordinates $(z,\bar{z})$  in a neighborhood $U_i$ of $x_i$
are  such that 
$$z(x_i)=0,$$ 
and $U_i$ itself is a coordinate disc of the coordinate radius $\epsilon$. 

In $U_i$, the function $F$ may be written as
\begin{equation}
F=z^{n_i} e^{g_i(z)}e^{-\phi_{U_i}}
\end{equation}
where  $n_i\in{\mathbb{N}}$ is  the degree of the zero at $x_i$, and $g_i(z)$ is holomorphic. 
The value of $n_i$ can be found by integrating the $1$-form 
\begin{equation}
\frac{\partial_{z} F}{F}dz = n_i\frac{dz}{z}+\partial_z\left(g_i(z)-\phi_{U_i}\right)dz,
\end{equation}
along the boundary $\partial U_i$ of the disc, namely 
\begin{equation}
{\rm lim}_{\epsilon\rightarrow 0}  \int_{\partial U_i}  \frac{\partial_z F}{F}dz = 2\pi i n_i.  
\end{equation}
The integrant can be written in a covariant way 
$$ \frac{\partial_z F}{F}dz=  \frac{dF^{(1,0)}}{F}, $$
defined in the coordinate invariant way in $S\setminus\{x_1,...,x_k\}$. 

On the other hand, if we repeat the construction for every zero $x_i$, $i=1,...,k$, and consider 
$\epsilon$ sufficiently small such that the discs do not intersect any other, then
\begin{equation}\label{sum}
\sum_{i=1}^k \int_{\partial U_i}   \frac{dF^{(1,0)}}{F} \ =\ -\int_{S\setminus \bigcup_i  U_i} d\left( \frac{dF^{(1,0)}}{F}  \right). 
\end{equation}
Notice, that in the domain of the integration on the right hand side $F$ nowhere vanishes. 

Now,  
\begin{equation}
d\left(  \frac{\partial_z F}{F}dz \right)  =   \frac{F\partial_{\bar z}\partial_zF-\partial_{\bar z}F\partial_zF}{F^2} d{\bar z}\wedge dz  =  
-d\left(  \frac{\partial_{\bar z} F}{F}d{\bar z} \right)=3d\left(\omega_{\bar z}d{\bar z}\right),
\end{equation}
where the emergence of the $1$-form $\omega_{\bar z}d{\bar z}$ is due to eq. (\ref{delta+3pi'}).   Therefore we can rewrite the eq. (\ref{sum}) as
\begin{equation}\label{sum'}
\sum_{i=1}^k \int_{\partial U_i}   \frac{dF^{(1,0)}}{F} \ =\ - 3\int_{S\setminus \bigcup_i  U_i} d\left(\omega^{(0,1)}\right).
\end{equation}
The advantage of expressing the right hand side by $\omega$ is, that  the integrant $d\left(\omega^{(0,1)}\right)$ is an exact $2$-form defined on 
the entire $S$ (including the points $x_1,...,x_k$). Hence
\begin{equation}
2\pi i\sum_i  n_i = {\rm lim}_{\epsilon\rightarrow 0}  \sum_{i=1}^k \int_{\partial U_i}   \frac{dF^{(1,0)}}{F} =-3 {\rm lim}_{\epsilon\rightarrow 0} \int_{S\setminus \bigcup_i  U_i} d\left(\omega^{(0,1)}\right) =-3\int_S d\left(\omega^{(0,1)}\right)=0.
\end{equation}

In summary, assuming that the function $F$ has isolated zeros we have come to the opposite conclusion that 
 it has  no zeros. It completes the proof. 

\medskip

\noindent{\bf Remark 2}  In the case  $S$ is simply connected itself (that is $S=S_2$) the functions
$\phi_U$ and $f_U$ are defined globally on $S$, hence we can drop the suffix $U$
$$\phi:=\phi_U, \ \ \ \ \ f:=f_U.$$
Moreover, as an entire  holomorphic function on $S$, 
$$f=f_0={\rm const}.$$
If we apply the eq.  (\ref{f_U}) to the scalar  $\Psi_2$ of a solution to the NHG, we find
$$\Psi_2=f_0e^{-\phi}, \ \ \ \ {\rm where}\ \ \ \  \partial_{\bar z}\phi=3\omega_{\bar z} . $$ 
In the $\Lambda=0$ case that equation was derived in \cite{LPextremal}.  
 
\section{The emergence of the type D equation} 
We go back now  to the NHG equation (\ref{NHG2d}) and properties of 
the (semi) scalar $\Psi_2$   (\ref{Psi2},\ref{Omega}) of a solution 
$(g_{AB}, \omega_A)$. It follows from  Proposition 1
and  Lemma 1 that either 
\begin{equation}\label{Psi0} \Psi_2=0 \end{equation}  
identically on $S$, or 
\begin{equation}\label{Psinot0}
 \Psi_2(x) \not= 0, \ \ \ \ \ {\rm for\ every}\ \ \ \ x\in S. 
\end{equation}   


In case (\ref{Psi0}) we have
$$d\omega = 0, \ \ \ \ \ K=\frac{\Lambda}{3}.$$
Since this solution  is static, according to \cite{CRT1}
$$\omega_A=0.$$  
But then the NHG equation implies
$$K=\Lambda.$$
Hence, the only solution is 
$$K=0=\Lambda$$
defined on a torus. 
 
Then, for the rest of this section  consider case (\ref{Psinot0}). The eq. (\ref{deltaPsi}) is equivalent to 
 \begin{equation}\label{deltaPsiroot}
\bar{m}^A\left(\nabla_A - \omega_A\right)\left(\Psi_2\right)^{-\frac{1}{3}}\ =\ 0.  
\end{equation}  
A function $\left(\Psi_2\right)^{-\frac{1}{3}}$ is defined up to rescaling by cubic 
roots of $1$ and unless  genus$(S)=0$, an existence of continues 
$\left(\Psi_2\right)^{-\frac{1}{3}}$ a priori is not guaranteed. However, 
given $g, \omega$ and the corresponding  nowhere vanishing $\Psi_2$, there is always a covering space 
$$\tilde{S}\rightarrow S$$ 
such that for the pullback $\tilde{g}, \tilde{\omega}$ and $\tilde{\Psi}_2$ 
there is a continues function  $\left(\tilde{\Psi}_2\right)^{-\frac{1}{3}}$. Moreover,
$${\rm genus}(\tilde{S})\geq  {\rm genus}(S),$$ 
and in particular, if  $S=T^2$, then $\tilde{S}$ is also $T^2$. 
That makes cases 1-3 of Sec. II preserved by going (or not) to the covering $\tilde{S}$.  
Hence, given $(g_{AB}), \omega_A$ and the corresponding $\Psi_2$, we  choose 
a suitable covering and drop the tildes. (Actually, it turns out that the resulting $\Psi_2$
is continues on the original $S$, without the need of covering by $\tilde{S}$.)

The inverse cubic root  in equation (\ref{deltaPsiroot})  reminds of the Type D equation
defined in \cite{DLP1,DKLS} that  also is an integrability condition for the NHG equation. 
The following observation can serve as an independent proof of that relation
between the equations:  
 \medskip

\noindent{\bf Lemma 2}  {\it  The following identity is true for arbitrary $f\in C^2(S)$:
 \begin{align}
&\bar{m}^B\bar{m}^A\nabla_B\nabla_A f = \left(\bar{m}^B\bar{m}^A\left( \nabla_{(A}\omega_{B)} + \omega_A\omega_B - \frac{1}{2}Kg_{AB} + \frac{1}{2}\Lambda g_{AB} \right)\right) f\nonumber\\   
&+\bar{m}^B\nabla_B\left(\bar{m}^A\left(\nabla_A - \omega_A\right)f\right)+ \left( \bar m^A \omega_A - \bar m^A \left(\nabla_A \bar m^C \right) m_C \right)\left( \bar m^B \left( \nabla_B - \omega_B \right)f\right)
\end{align}
}

The proof is just a straightforward calculation of the first term on the right hand side, and noticing, that 
the terms proportional to $K$, and to $\Lambda$ in fact do not contribute. They are there
only for the relation with the NHG equation.    
\medskip 
 
The conclusion from Lemma 2 is, that if the NHG equation is satisfied by $(g_{AB}, \omega_A)$, then the corresponding $\Psi_2$
satisfies an equation 
\begin{equation}
\bar{m}^B\bar{m}^A\nabla_B\nabla_A \left(\Psi_2\right)^{-\frac{1}{3}} =   0.
\end{equation}
We call it  Type D equation. The subject of the equation  is $(g_{AB}, \omega_A)$ \cite{DLP1}. 

This equation has an equivalent holomorphic formulation.  It can be shown \cite{DKLS}, that  a function $F$ satisfies the equation   
\begin{equation}\label{ddf}
\bar{m}^B\bar{m}^A\nabla_B\nabla_A F =   0
\end{equation}
if and only if  the vector field 
\begin{equation}\label{holv}
 \left( f_{,B}g^{BA}\right)^{(1,0)} 
\end{equation}
 is a holomorphic vector field on $S$.

\section{The general solution to NHG equation for genus$(S)>0$ }
The observation that (\ref{holv}) is a holomorphic vector field defined globally on $S$,   is useful in the derivation of a general solution  of (\ref{ddf}) 
for  ${\rm genus}(S)>0$.  The only solutions are \cite{DKLS} 
\begin{equation}
F = {\rm const}. 
\end{equation}
In a genus$(S)>1$ case that result  follows immediately from the non-existence of non-trivial holomorphic vectors. In the case
of genus$(S)=1$ (torus), the dimension of the space of the holomorphic vectors is $1$, however none of them is   the gradient
of a function. 

Hence, the almost scalar $\Psi_2$ corresponding to $(g_{AB},\omega_A)$ that solves 
the NHG equation (\ref{NHG2d}) on $S$ of genus$>1$  in case (\ref{Psinot0})
is
$$\Psi_2={\rm const}\not=0. $$
We could  integrate  the  definition of $\Psi_2$ (\ref{Psi2}) with respect to $g_{AB}$ and $\omega_A$. 
However,  the eq. (\ref{deltaPsiroot})  is more powerful:
$$ 0\ =\ \bar{m}^A(\nabla_A + 3\omega_A)\Psi_2\ =\ 3\bar{m}^A\omega_A\Psi_2.$$
 That implies 
\begin{equation}
\omega^A\ =\ 0.
\end{equation}
Substituting that result into the original NHG equation (\ref{NHG2d}) we find, that
\begin{equation}
K=\Lambda. 
\end{equation} 

The careful reader notices that we still have to go back to the original 
$2$-manifold $S$ that was covered by $\tilde{S}$ for the continuity
of the cubic root of $\Psi_2$. However, the resulting $\Psi_2$ on the
original $S$ is constant, hence its cubic root is continues on $S$. 
\medskip

\section{Summary}  The most important conclusion coming from this work is:

\medskip

\noindent{\bf Theorem}{\it\ The only solutions to the Near Horizon Geometry equation (\ref{NHG2d}) with
a cosmological constant $\Lambda$  on a $2$-dimensional, compact, orientable  manifold
of 
$${\rm genus} >  0$$
are pairs $(g_{AB},\omega_A)$ such that 
$$K=\Lambda, \ \ \ \ \ \ \ \omega_A=0.$$
}

The theorem  relies on earlier results about:  $(i)$ the solutions to the Type D equation \cite{DLP1, DKLS}, 
and $(i)$ the static solutions of the NHG equation \cite{CRT1} - we needed
those to complete the $\Psi_2=0$ case. 

The essentially new part that fills in the gap is Lemma 1 proved in the current paper which ensures, that 
if $(g_{AB},\omega_A)$ satisfy the NHG equation on a compact connected surface $S$, then 
the invariant $K-\frac{\Lambda}{3} +i\Omega$ is either everywhere or nowhere, zero on $S$.

Due to that result, the invariant satisfies the Type D equation at every point of an orientable compact $S$, 
that is
\begin{equation}\label{TypeD}
\bar{m}^A\bar{m}^B\nabla_A\nabla_B(K-\frac{\Lambda}{3} +i\Omega)^{-\frac{1}{3}}\ = 0,
\end{equation}
where at the end of the day it turns out that   the cubic root function is continuous on $S$.

Consequently, the equation 
\begin{equation}\label{delta+omegapsi}
\bar{m}^A \left(\nabla_A +3\omega_A \right) (K-\frac{\Lambda}{3} + i\Omega) = 0,
\end{equation}
is a generalization to $\Lambda\not=0$ of the equation 
\begin{equation}
\bar{m}^A \left(\nabla_A +3\omega_A \right) (K + i\Omega) = 0,
\end{equation}
used before \cite{LPextremal,LRS} in the $\Lambda=0$ case. 
\medskip
It makes Lemma 1 and the Type D equation applicable also to still unsolved  $S=S_2$ case  and the NHG equation with the cosmological constant $\Lambda<0$.

Finally, our results extend to non-orientable $2$-dimensional compact manifolds
due to the existence of  orientable double coverings.   Given $S, g_{AB}, \omega_A$
and an orientable covering 
$$ \check{S}\rightarrow S$$
we just pull back the data and obtain $\check{S}, \check{g}_{AB}, \check{\omega}_A$.
The NHG equation is consistent with that map. In this way we  extend our results to
non-orientable manifolds in the following way: 
 \medskip
 
\noindent{\bf Theorem'}{\it\ The only solutions of the Near Horizon Geometry equation (\ref{NHG2d}) with
a cosmological constant $\Lambda$  on a $2$-dimensional, compact manifold
$S$ of the Euler characteristics  
$$\chi_E(S) \leq 0$$
are pairs $(g_{AB},\omega_A)$ such that 
$$K=\Lambda, \ \ \ \ \ \ \ \omega_A=0.$$
}
\medskip
 
 \noindent{\bf Lemma 1'}{\it\ If $S$ is a compact $2$-manifold and $g_{AB}, \omega_A$ satisfy the NHG equation
 (\ref{NHG2d}) then at every point $x\in S$,
$$ K(x) \not=0 \ \ \ {\rm or}\ \ \ d\omega_{|_x}\not=0 . $$ 
} 
 
The   complex almost invariant $K-\frac{\Lambda}{3}+i\Omega$ is not invariant anymore.
It is defined locally, upon the choice of the null frame $m^A$, however in addition to the transformations 
(\ref{m'}) we need also
 $$m'^A = e^{i\alpha}\bar{m}^A.$$
 They have to be accompanied by
 $$K'=K, \ \ \ \  \omega'_A=\omega_A,\ \ \ \ \Omega'=-\Omega.$$
Then the Type D equation  defined locally, passes consistently 
from an orientable neighborhood to another overlapping orientable 
neighborhood.  The same rule applies to equation (\ref{delta+omegapsi}). 
 \medskip

\noindent{\bf Acknowledgements:}
This work was partially supported by the Polish National Science Centre grant No. 2015/17/B/ST2/02871.

\end{document}